\documentstyle[epsf,aps]{revtex}

\begin{document}

\title{The nonlinear evolution of de Sitter space instabilities}

\author{Jens C. Niemeyer} 
\address{Enrico Fermi Institute, University of Chicago, Chicago, Illinois 60637} 
\author{Raphael Bousso}
\address{Department of Physics, Stanford University, Stanford,
California 94305} 

\maketitle

\begin{abstract}
  
We investigate the quantum evolution of large black holes that
nucleate spontaneously in de~Sitter space. By numerical computation in
the s-wave and one-loop approximations, we verify claims that
such black holes can initially ``anti-evaporate'' instead of
shrink. We show, however, that this is a transitory effect. It is
followed by an evaporating phase, which we are able to trace until the
black holes are small enough to be treated as Schwarzschild.  Under
generic perturbations, the nucleated geometry is shown to decay into a
ring of de~Sitter regions connected by evaporating black holes. This
confirms that de~Sitter space is globally unstable and fragments into
disconnected daughter universes.

\end{abstract}

\section{Introduction and Summary}

\subsection{Anti-evaporation and proliferation}

Schwarzschild-de~Sitter black holes have unusual quantum properties
and instabilities%
~\cite{GibHaw77a,GinPer83,BouHaw96,BouHaw97b,NojOdi98b,%
EliNoj99,Bou98,Bou99}. They are of cosmological interest because they
can be produced during inflation~\cite{BouHaw95}. They are also
theoretically significant because they change the global structure of
de~Sitter space~\cite{Bou98,Bou99} fundamentally.

Unlike their asymptotically flat cousins, Schwarzschild-de~Sitter
black holes are surrounded by a cosmological horizon. This limits
their size and puts them in a thermal bath. Their temperature is
always larger than that of the cosmological
horizon~\cite{GibHaw77a,BouHaw96}.  Nevertheless, Hawking and one of
the authors argued in Ref.~\cite{BouHaw97b} that some black holes
accrete so much quantum radiation that they will grow, or
`anti-evaporate,' instead of evaporating (see
also~\cite{NojOdi98b,EliNoj99}).

In the maximal solution, the black hole and cosmological horizon are
of equal size, and the spatial geometry will be $S^1 \times S^2$, with
constant two-sphere radius. This geometry can nucleate semiclassically
in de~Sitter space through a gravitational tunneling process.  Its
evolution is unstable to the formation of $n$ de~Sitter regions,
distributed around the $S^1$ and connected by $n$ black hole
throats~\cite{Bou98}. This configuration may be visualized as a
`doughnut' with $n$ `wobbles,' whose minima and maxima represent the
black hole and cosmological horizons, respectively. For $n=1$, this
reduces to the ordinary, submaximal Schwarzschild-de~Sitter solution.
For higher $n$, it resembles a ring-like sequence of $n$
Schwarzschild-de~Sitter solutions. As the de~Sitter regions form
between the cosmological horizons, the black holes are expected to
evaporate and shrink.  When they finally disappear, the doughnut is
pinched in $n$ places, leaving $n$ disconnected pieces. This
corresponds to the fragmentation of space into $n$ large daughter
universes~\cite{Bou98}.

Thus de~Sitter space is globally unstable. Locally, however, the
daughter universes are indistinguishable from de~Sitter.  Therefore,
they will harbour more maximal black holes, whose evaporation can lead
to further fragmentation. Iteratively, an unbounded number of
disconnected de~Sitter universes is produced: de~Sitter space
proliferates.  This drastic conclusion depends crucially on the
nonlinear evolution of the instabilities of the maximal
Schwarzschild-de~Sitter solution (also referred to as the `Nariai'
solution below).

\subsection{Analytical shortcomings}

In Refs.~\cite{BouHaw97b} and~\cite{Bou98,Bou99}, the anti-evaporation
and proliferation effects were derived analytically, using a number of
approximations. The stability of the Nariai geometry was studied using
linear perturbation theory. The evolution of the horizons was found
only for very early times, as a power series in the time variable.
This led to the discovery of anti-evaporation for a class of
perturbations. But a power series can prove only {\em initial}
anti-evaporation. It cannot answer an important question: Will such
black holes continue to grow, asymptotically approaching the maximal
(Nariai) solution? Or will their growth eventually stall, and
evaporation set in after all? This problem is highlighted by recent
claims that anti-evaporation may permanently stabilize black holes
nucleated during inflation\footnote{We point out, however, that even
stable black holes nucleated semiclassically during inflation would be
diluted by the cosmological expansion. Typically, they will lie
outside the current horizon and will not be
observable~\cite{BouHaw96,Bou96b}.}~\cite{NojOdi98b}.

At least for Euclidean boundary conditions, it was shown that the
black holes do evaporate at late times~\cite{BouHaw97b}. In fact,
arguments might be made that the evaporating mode is attractive at
late times for all initial conditions~\cite{Bou99}. In any case,
however, the intermediate period containing the putative transition
between anti-evaporation and evaporation is beyond the reach of the
analytical methods employed. Can there be a smooth transition, and how
does it proceed? To understand this crucial phase of the black hole
evolution, it should be explored numerically.

If anti-evaporation is only temporary, then how can we be sure that
the evaporating phase, once entered, continues until the black hole
disappears?  This question is particularly important in the context of
multiple black hole formation and proliferation~\cite{Bou98,Bou99}.
The fragmentation of space is possible only if all black holes become
small and disappear like Schwarzschild black holes.

To show that black holes really evaporate completely, it is necessary
to trace the backreaction at least until they are significantly
smaller (perhaps by a factor of 10)
than the cosmological horizon. Then they can be assumed to behave like
Schwarzschild black holes, which presumably radiate all of their mass
away. But linear perturbation theory requires that the small
two-spheres inside the black hole and the large two-spheres between
the cosmological horizons still be of nearly equal size. Thus, the
analytic `late-time' solution, showing black hole evaporation, is not
really valid for arbitrarily late times. A full, non-linear numerical
calculation is needed to interpolate to the Schwarzschild regime.

\subsection{Numerical analysis}

In Sec.~\ref{sec-antievap} we consider $n=1$ perturbations of
the Nariai geometry, which can be parametrized by their initial
amplitude, $\epsilon$, and phase, $\vartheta$.  Depending on
$\vartheta$, the black holes undergo initial periods of
anti-evaporation or evaporation.  We demonstrate the transitory nature
of anti-evaporation for neutral black holes by showing that
evaporation sets in at sufficiently late times, independently of
$\vartheta$. The evaporation rate is compared to an approximate
solution of the linear equations in order to verify that this solution
is an attractor. Beyond the linear regime, the black holes are found
to shrink to a size much smaller than the cosmological horizon before
our computation breaks down. Then they live in approximately flat
space, and should simply continue to evaporate until they disappear.

In Sec.~\ref{sec-charged} we show that maximal black holes become
stable if their charge exceeds a critical value, in agreement with
thermodynamic expectations. This confirms analytic arguments given in
Ref.~\cite{Bou99}. In particular, we show that supercritical black
holes, independently of their initial behavior, always end up
anti-evaporating.  This is the condition for a novel type of
daughter-universe production proposed in Ref.~\cite{Bou99}.

In Sec.~\ref{sec-proliferation} we show that a higher mode ($n>1$)
perturbation behaves as predicted by analytical arguments: First it
oscillates, then it freezes out, forming black hole interiors and
growing de~Sitter regions. Depending on initial conditions, a period
of anti-evaporation may follow. Finally, the black holes evaporate and
become much smaller than the cosmological horizons. 

\section{Action and perturbations}
\label{sec-model}

\subsection{Including back-reaction}

The four-dimensional Lorentzian Einstein-Hilbert action with a
cosmological constant, $\Lambda$, and a Maxwell field, $F_{\mu\nu}$,
is given by:
\begin{equation}
S = \frac{1}{16 \pi} \int d^4\!x\, (-g^{{\rm IV}})^{1/2} \left[
 R^{{\rm IV}} - 2 \Lambda - F_{\mu\nu} F^{\mu\nu} \right],
\label{eq-action-4D}
\end{equation}
where $R^{{\rm IV}}$ and $g^{{\rm IV}}$ are the four-dimensional Ricci
scalar and metric determinant.

Restricting to spherically symmetric fields and quantum fluctuations,
the metric may be written as
\begin{equation}
ds^2 = e^{2\rho} \left( -dt^2 + dx^2 \right) + e^{-2\phi} d\Omega^2,
\label{eq-ssans}
\end{equation}
where $x$ is the coordinate on the $S^1$, with period $2\pi$.  Using
this ansatz, and the on-shell condition for magnetic fields,
\begin{equation}
F_{\mu\nu} F^{\mu\nu} = 2 Q^2 e^{4\phi},
\end{equation}
the angular coordinates and the Maxwell field can be integrated out in
Eq.~(\ref{eq-action-4D}), which reduces the action to
\begin{equation}
S = \frac{1}{16\pi} \int d^2\!x\, (-g)^{1/2} e^{-2\phi} \left[ R + 2
 (\nabla \phi)^2 + 2 e^{2\phi} - 2 \Lambda - 2 Q^2 e^{4\phi} \right],
\end{equation}

In order to include back-reaction effects, a Polyakov term, which
arises in the one-loop effective action of a two-dimensional scalar
field, will be included.  (It would be preferable to work with
dilaton-coupled
scalars~\cite{EliNaf94,MukWip94,ChiSii97,BouHaw97a,NojOdi97,%
NojOdi98e,Ich97,KumLie97,Dow98,GusZel99,LomMaz99,BalFab99}, or even
better, with a four-dimensional effective action reduced to
two-dimensions.  This would complicate the numerical computation
enormously and will not be attempted here.  For small perturbations,
it has been shown that extra terms from dilaton-coupling do not affect
results~\cite{BouHaw97b,Bou99}.  Thus one would not expect qualitative
changes even for black holes noticeably smaller than the cosmological
horizon.  In order to corroborate our results for nearly-Schwarzschild
black holes, however, a full four-dimensional treatment would be
desirable in future work.)  In the large $N$ limit, the contribution
from the quantum fluctuations of the scalars dominates over that from
the metric fluctuations.  In order for quantum corrections to be
small, one should take $ N \Lambda \ll 1 $.  One can obtain a local
form of this action by introducing an independent scalar field $Z$
which mimics the trace anomaly~\cite{BouHaw97b}.  The on-shell
equivalence of the equations of motion can be seen by choosing a
conformal gauge for the two-dimensional metric, as in Eq.~(2). Thus
the action of the one-loop model will be given by:
\begin{equation}
S = \frac{1}{16\pi} \int d^2\!x\, (-g)^{1/2} \left[
 \left( e^{-2\phi} + \frac{ N }{3} Z \right) R
 - \frac{ N }{6} \left( \nabla Z
 \right)^2 + 2 + 2 e^{-2\phi} \left( \nabla \phi \right)^2 - 2
 e^{-2\phi} \Lambda -2 Q^2 e^{2\phi} \right].
\end{equation}

\subsection{Equations of motion}

Differentiation with respect to $t$ ($x$) will be denoted by an
overdot (a prime). For any functions $f$ and $g$, define:
\begin{equation}
\partial f\,\partial g  \equiv - \dot{f} \dot{g} + f' g',\ \ \ \
\partial^2 g \equiv - \ddot{g} + g'',
\end{equation}
\begin{equation}
\delta f\,\delta g \equiv \dot{f} \dot{g} + f' g',\ \ \ \ \delta^2 g
\equiv \ddot{g} + g''.
\end{equation}
Variation with respect to $\rho$, $\phi$ and $Z$ yields the following
equations of motion:
\begin{equation}
- \partial^2 \phi + 2 (\partial \phi)^2 + \frac{N}{6} e^{2\phi}
\partial^2 Z + e^{2\rho+2\phi} \left( \Lambda e^{-2\phi} + Q^2
e^{2\phi} - 1 \right) = 0;
\label{eq-m-rho}
\end{equation}
\begin{equation}
\partial^2 \rho - \partial^2 \phi + (\partial \phi)^2 + \Lambda
e^{2\rho} - Q^2 e^{2\rho+4\phi} = 0;
\label{eq-m-phi}
\end{equation}
\begin{equation}
\partial^2 Z - 2 \partial^2 \rho = 0.
\label{eq-m-Z}
\end{equation}
The constraint equations are:
\begin{equation}
 \left( \delta^2 \phi - 2 \delta\phi\,\delta\rho \right) -
(\delta\phi)^2 = \frac{N}{12} e^{2\phi} \left[ (\delta Z)^2 + 2
\delta^2 Z - 4 \delta Z \delta\rho \right];
\label{eq-c1}
\end{equation}
\begin{equation}
 \left( \dot{\phi}' - \dot{\rho} \phi' - \rho' \dot{\phi} \right) -
\dot{\phi} \phi' = \frac{N}{12} e^{2\phi} \left[ \dot{Z} Z' + 2
\dot{Z}' - 2 \left( \dot{\rho} Z' + \rho' \dot{Z} \right) \right].
\label{eq-c2}
\end{equation}
From Eq.~(\ref{eq-m-Z}), it follows that $Z = 2\rho + \eta$, where
$\eta$ satisfies $\partial^2 \eta = 0$. The remaining freedom in
$\eta$ can be used to satisfy the constraint equations for any choice
of $\rho$, $ \dot{\rho} $, $\phi$ and $\dot{\phi}$ on an initial
spacelike section~\cite{BouHaw97b}.

\subsection{Metric and horizon perturbation}

Maximal (Nariai) black holes nucleate semiclassically in de~Sitter
space. This process is mediated by gravitational instantons and has
been described in detail in Refs.~\cite{GinPer83,BouHaw95,BouHaw96}.
Here we are interested not in the nucleation, but in the further
evolution of the Nariai solution.  Its metric is given by
\begin{equation}
e^{2\rho} = \frac{1}{A} \frac{1}{\cos^2\! t}, \ \ \ e^{2\phi} = B,
\label{eq-CN-metric2}
\end{equation}
where $B$ is given by the cubic equation
\begin{equation} % from Qw2
B \left[ 1 - Q^2 B \left( 1 + \frac{N}{3} B \right) \right] = \left[ 1
- \frac{N}{3} B \right] \Lambda;
\label{eq-B}
\end{equation}
the physical solution is the one that limits to $B=\Lambda$ for
$N=Q=0$). $A$ is given by
\begin{equation}
A = \Lambda - Q^2 B^2.
\end{equation}

Quantum fluctuations will perturb this solution, so that the
two-sphere radius, $e^{-\phi}$, will vary slightly along the
one-sphere coordinate, $x$. Decomposition into Fourier modes on the
$S^1$ yields the perturbation ansatz
\begin{equation}
e^{2\phi} =  \Lambda_2 \left[ 1 + 2 \epsilon \sum_n \left( \sigma_n(t)
 \cos nx + \tilde{\sigma}_n(t) \sin nx \right) \right],
\label{eq-fullpert-phi}
\end{equation}
where $\epsilon$ is taken to be small. This will be referred to as the
{\em metric perturbation}, characterized by the $\sigma_n$,
$\tilde{\sigma}_n$ at the time $t=0$.

More generally, one should also consider perturbations of the time
derivative of the two-sphere radius, expressed by $\dot \sigma_n$. As
in Ref.~\cite{Bou99}, we parameterize the initial conditions for the
perturbation as
\begin{equation}
\sigma_n(0) = \sin \vartheta \qquad , \qquad \dot \sigma_n(0) = \cos
\vartheta\,\,,
\end{equation}
where $\vartheta$ represents the phase of the initial perturbation.

Based on linear perturbation theory, one would expect such
perturbations to lead to a classical, as well as a quantum
instability~\cite{GinPer83,BouHaw95,BouHaw96,BouHaw97b,Bou98,Bou99}.
Classically, the regions on the $S^1$ where the two-spheres are
smaller than the Nariai value should collapse to form black hole
interiors. The larger two-spheres, on the other hand, should grow
exponentially, developing into asymptotic de~Sitter regions. If the
first mode dominates, this simply leads to a nearly-maximal
Reissner-Nordstr\"om-de~Sitter solution. But if higher modes are
strongly excited, a whole sequence of Reissner-Nordstr\"om-de~Sitter
solutions can develop around the same one-sphere. This may be thought
of as a necklace of de~Sitter regions, strung together by black hole
throats.

The expected quantum instability is the evaporation of these black
holes.  Actually, we will show below that black holes of sufficient
charge are stable.  But all other black holes would be expected to
evaporate and get smaller, since their temperature is higher than that
of the surrounding cosmological horizon.

The black hole evolution can be calculated numerically using
Eqs.~(\ref{eq-m-rho})--(\ref{eq-m-Z}).  It is important to stress that
the black hole and cosmological horizons do not in general correspond
to the minimal and maximal two-spheres along the $S^1$, although such
a slicing can always be found.  In general, one must first find the
positions of the horizons on the one-sphere; this can be done by
finding the points where the gradient of the two-sphere size is
null~\cite{BouHaw97b}. The two-sphere sizes at those locations give
the size of the black hole. The black hole evolution can be monitored
by following the horizon location and plotting the horizon size vs.\
time.  It will be convenient to define the horizon perturbation
$\delta$, for a black hole located at $x_{\rm b}$, by
\begin{equation}
r_{{\rm b}}(t)^{-2} = e^{2\phi[t, x_{{\rm b}}(t)]} = B \left[ 1 + 2
\epsilon \delta(t) \right].
\label{eq-rb}
\end{equation}
Thus $\delta$ corresponds to the fractional difference between the
current black hole size and the size of a maximal black hole of equal
charge.  Evaporation corresponds to increasing values of $\delta$.

\begin{figure}
\epsfxsize=0.45\textwidth
\epsfbox{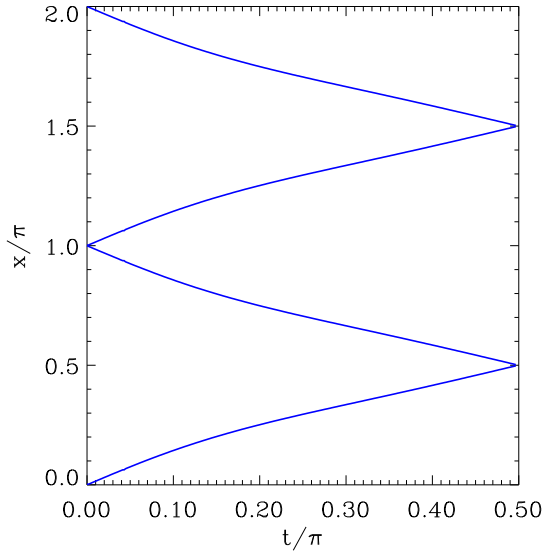}
\caption{\label{fig-horizons} Location of the cosmological and black
hole horizons on the $S^1$ coordinate for the perturbation mode
$n=1$.}
\end{figure}

Because the $S^1$ expands exponentially, the black hole and
cosmological horizons of a Reissner-Nordstr\"om-de~Sitter solution
appear to move ever more closely together in comoving $S^1$
coordinates (see Fig.~\ref{fig-horizons}).

\subsection{Numerical technique}
\label{numerics}

The equations of motion (\ref{eq-m-phi}) -- (\ref{eq-m-Z}) were solved
numerically using the standard method of characteristics for second
order quasi-linear hyperbolic equations~\cite{smith85}.  For this
purpose, the auxiliary field $Z$ can be  eliminated by inserting
Eq.~(\ref{eq-m-Z}) into Eq.~(\ref{eq-m-phi}) and the remaining
equations can be rewritten in manifestly hyperbolic form:
\begin{equation}
\partial^2 \phi = {\cal C} \left[ e^{2 \rho} \left(3 \Lambda - e^{2 \phi} (3
+ N \Lambda) + 3 e^{4\phi}Q^2 + e^{6\phi} N Q^2 \right) - (e^{2\phi}N
- 6)(\partial \phi)^2 \right]
\end{equation}
\begin{equation}
\partial^2 \rho = 3 {\cal C} \left[e^{2(\rho+\phi)} \left(2 e^{2\phi}Q^2
-1\right) + (\partial \phi)^2 \right]\,\,,
\end{equation}
where
\begin{equation}
{\cal C} = \frac{1}{3 - e^{2\phi}N}\,\,.
\end{equation}
The charateristic curves, along which the equations reduce to ordinary
differential equations, are $x_{\pm} = x \pm t$. Given the solution at
two neighboring points on non-identical characteristics and
approximating $dx_{\pm}$ by $\Delta x_{\pm}$, the solution at the
intersection of the characteristic curves going through these points
can be found by iteration. Assigning the initial conditions
Eqs.~(\ref{eq-CN-metric2}) -- (\ref{eq-fullpert-phi}) on a $t=0$
hypersurface thus allows the advancement of the solution along the
$x_{\pm}$-grid, employing periodic boundary conditions at $x = 0$ and
$x = 2 \pi$.

For most of our computations, we used a resolution of 10000 grid
points on $x$-hypersurfaces. The accuracy of the results was verified
by comparison with perturbative solutions \cite{BouHaw97b} and with
results obtained from an independent second-order finite difference
code for the same equations. With the exception of the very last time
step, the closest to the conformal time coordinate singularity,
excellent agreement of all solutions was found.

\section{Anti-evaporation and turnaround}
\label{sec-antievap}

\begin{figure}
\begin{minipage}{0.45\textwidth}
\epsfxsize=\textwidth \epsfbox{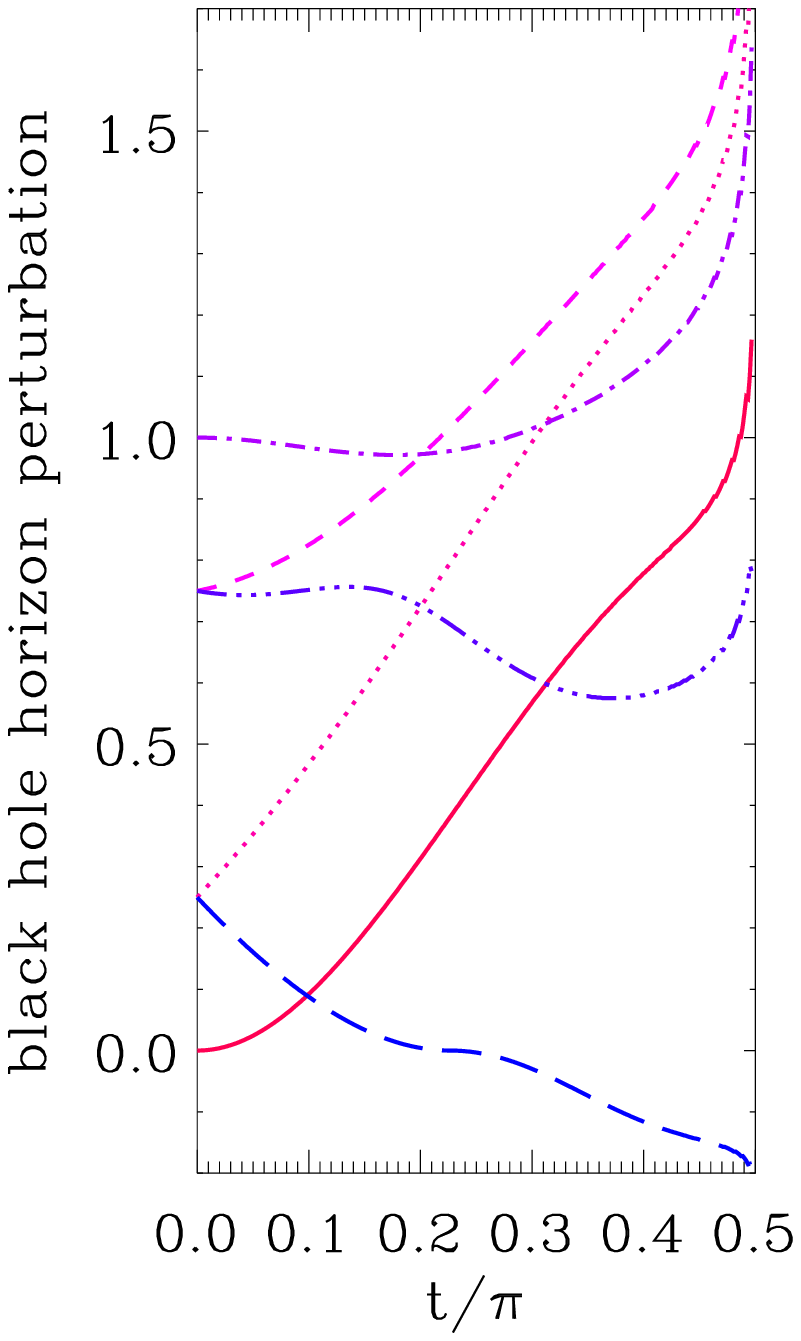}
\caption{\label{fig-6zw0per} Evolution of the black hole horizon
perturbation $\delta$  for different initial values of $\vartheta$:
$\vartheta = 0$ ({\sl solid}), $\vartheta =  \pi/6$ ({\sl dotted}),
$\vartheta = \pi/3$ ({\sl short dashed}), $\vartheta = \pi/2$ ({\sl
dashed-single dotted}), $\vartheta = 2 \pi/3$ ({\sl dashed-triple
dotted}), $\vartheta = 5 \pi/6$ ({\sl long dashed}) for $\Lambda =
0.1$, $N = 5$, and $w = 0$. The behavior is independent of the initial
amplitude $\epsilon$ as long as $\epsilon \ll 1$; all plots are for
$\epsilon = .0005$ unless otherwise indicated.}
\end{minipage}
\hspace{\fill}
\begin{minipage}{0.45\textwidth}
\epsfxsize=\textwidth 
\epsfbox{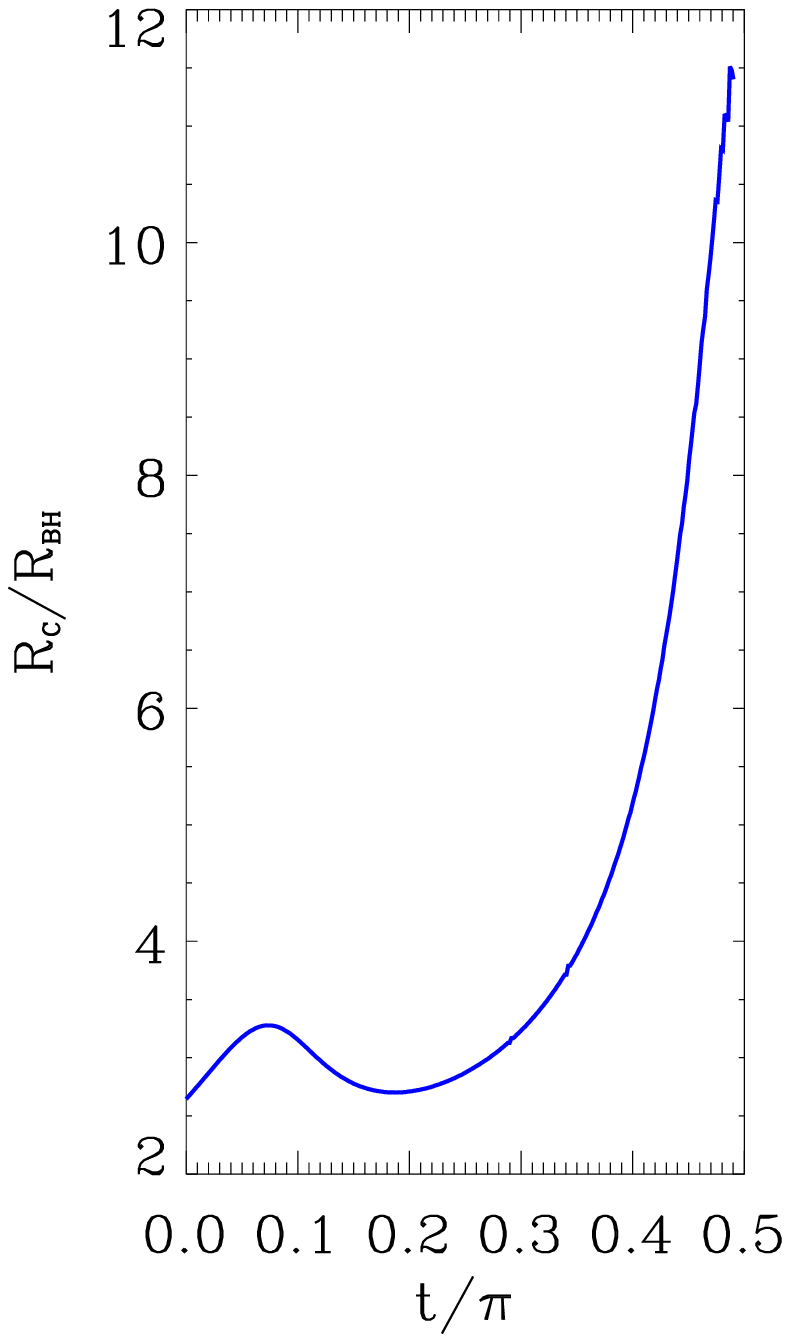}
\caption{\label{fig-schsch} Numerically one can show that perturbed
Nariai black holes evaporate until are significantly smaller than the
cosmological horizon. Then they can simply be viewed as living in
asymptotically flat space, so evaporation will continue. Model
parameters here are $\epsilon = 0.5$, $\vartheta = \pi/3$.  For this
run, an RST-type counterterm [24] was included in the
effective action in order to ensure that the black hole singularity
resides at $r=0$.}
\end{minipage}
\end{figure}

Anti-evaporation was first found in Ref.~\cite{BouHaw97b} for the
$\vartheta=\pi/2$, $n=1$ perturbation of the maximal (Nariai) neutral
black hole solution in de~Sitter space.  A power series approximation
showed that black holes formed by this perturbation will grow
initially.  A numerical calculation, however, allows us to probe the
black hole evolution beyond the range of validity of the power
series. Interestingly, this proves the anti-evaporation effect to be
transitory.  This is demonstrated by the long-dashed line in
Fig.~\ref{fig-6zw0per}. The horizon perturbation first decreases
quadratically, signaling anti-evaporation, in quantitative agreement
with the result of Ref.~\cite{BouHaw97b}. At a time $t \approx .2$,
however, it turns around and starts to increase. This means that the
black hole eventually stops growing. Instead, it starts to shrink in
size, corresponding to evaporation at late times.

For other initial phases, the behavior of the black hole can be quite
complicated, going through various periods of evaporation and
anti-evaporation (Fig.~\ref{fig-6zw0per}). The important result is,
however, that it always ends up evaporating at late times, as
conjectured in Ref.~\cite{Bou99}.  The numerical result for the final
evaporation rate agrees with the asymptotic solutions given there to
within better than one percent:
\begin{equation}
\delta \propto (\cos t)^{1-c_+},
\end{equation}
where $c_+$ is the larger root of
\begin{equation}
c(c+1) = 2 \sqrt{1 + \frac{2}{3} N \Lambda + \frac{1}{9} N^2
\Lambda^2}.
\end{equation}

An initial phase just under $\vartheta=\pi/2$ corresponds to a large
time-derivative of the metric perturbation directed opposite to the
perturbation.  In this case, the metric overshoots the Nariai value
and a cosmological horizon forms. The corresponding black hole is
located where the cosmological horizon would usually sit. Its
evaporation is mirrored by the growth of the cosmological horizon. In
Fig.~\ref{fig-6zw0per} this cosmological horizon shows up for
$\vartheta=5\pi/6$ as a negative value of $\delta$ which increases in
magnitude.

How can we be sure that the evaporation seen in the later stages of
Fig.~\ref{fig-6zw0per} is permanent?  By the nature of the conformal
time $t$, an infinite amount of proper time is contained in the final
step of the numerical calculation.  No matter how refined the step
size, this region cannot be resolved.  It is sufficient, however, to
trace the evolution until the black hole size differs significantly
from the size of the cosmological horizon.  This is precisely the
condition for non-linearity, so it cannot be verified analytically
using the power series for $\delta$. Our numerical calculation,
however, can reach this regime (see Fig.~\ref{fig-schsch}).  Once it
is so small, the black hole will be much hotter than the cosmological
radiation, and will no longer be influenced by it.  It will behave
like a Schwarzschild black hole, so we can trust that it will continue
to evaporate.

\begin{figure}
\begin{minipage}{0.45\textwidth}
\epsfxsize=\textwidth \epsfbox{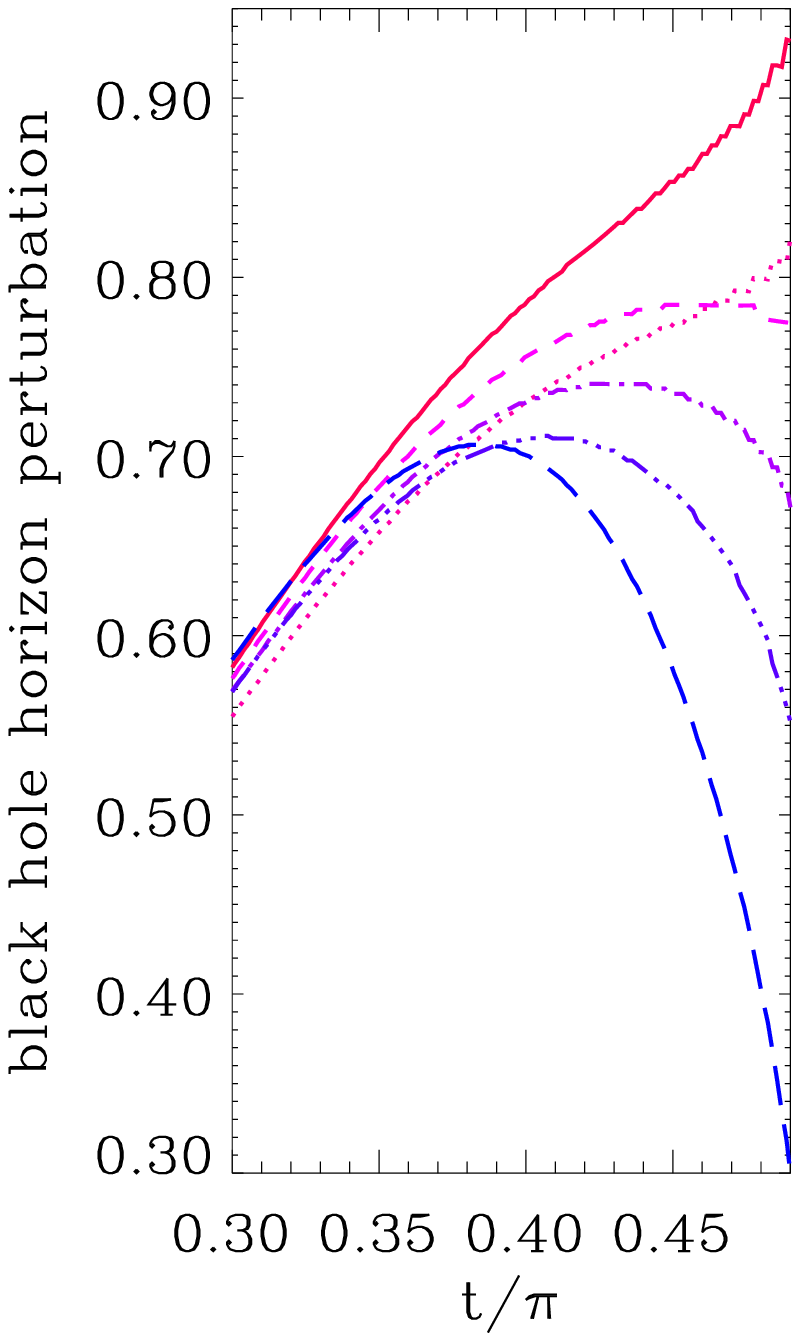}
\caption{\label{fig-z06Qw0per}Evolution of the horizon perturbation
$\delta$ for black holes with charges $Q^2 \Lambda = 0.195$ ({\sl
solid}), $Q^2 \Lambda = 0.215$ ({\sl dotted}), $Q^2 \Lambda = 0.235$
({\sl dashed}), $Q^2 \Lambda = 0.245$ ({\sl dashed-single dotted}), $Q^2
\Lambda = 0.255$ ({\sl dashed-triple dotted}), and $Q^2 \Lambda =
0.265$ ({\sl long dashed}). Model parameters were $\Lambda =
0.1$, $N = 5$, and $\vartheta = 0$.} 
\end{minipage}
\hspace{\fill}
\begin{minipage}{0.45\textwidth}
\epsfxsize=\textwidth \epsfbox{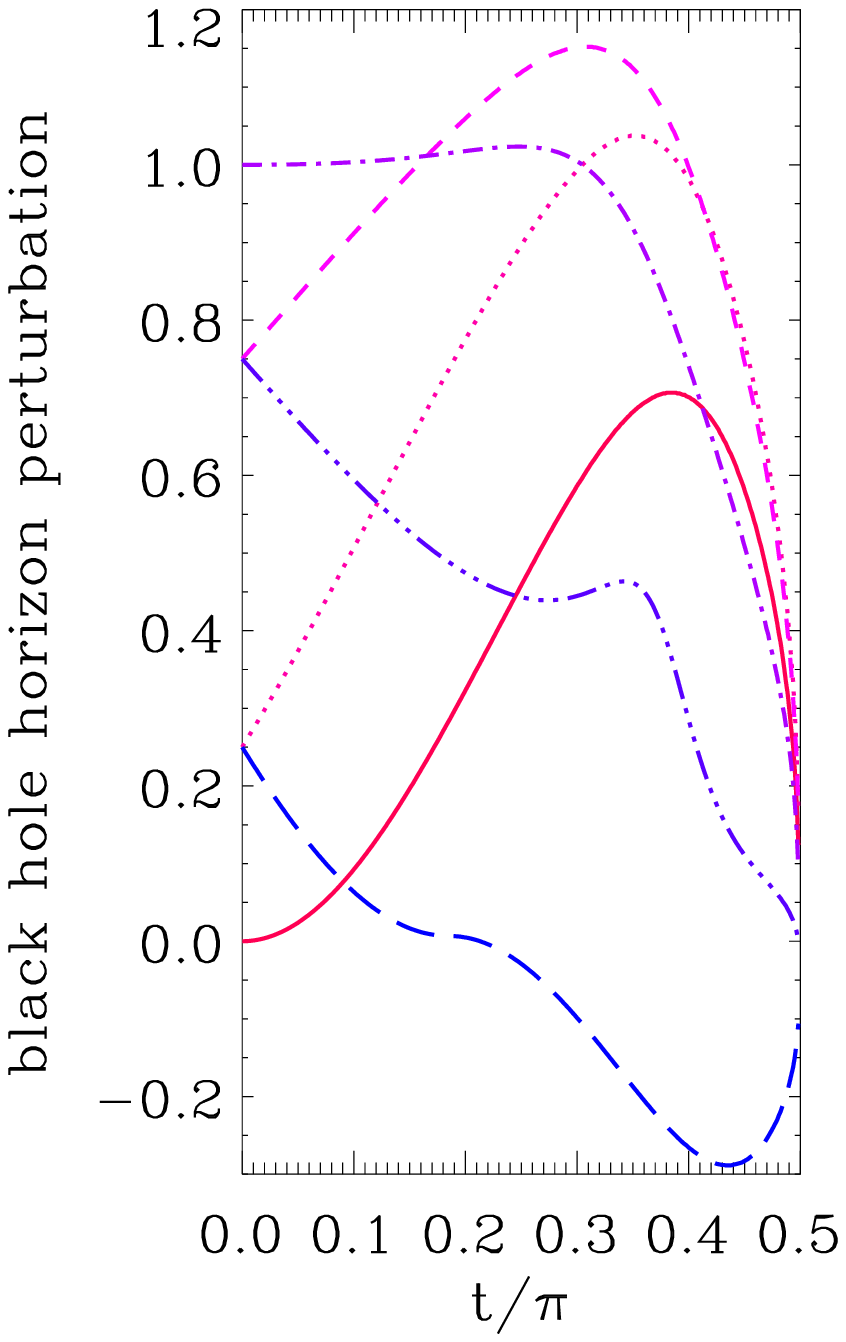}
\caption{\label{fig-6zw0Q265per} Evolution of the horizon perturbation
$\delta$ for slightly supercritically charged ($Q^2 \Lambda = 0.265$)
black holes with different initial values of $\vartheta$: $\vartheta =
0$ ({\sl solid}), $\vartheta = \pi/6$ ({\sl dotted}), $\vartheta =
\pi/3$ ({\sl dashed}), $\vartheta = \pi/2$ ({\sl dashed-single
dotted}), $\vartheta = 2 \pi/3$ ({\sl dashed-triple dotted}),
$\vartheta = 5 \pi/6$ ({\sl long dashed}); model parameter values as
in Fig.~\ref{fig-z06Qw0per}.  In this case the Nariai geometry is an
attractor.}
\end{minipage}
\end{figure}

\section{Stability of charged black holes}
\label{sec-charged}

Nariai solutions can be electrically or magnetically charged. In this
case, their spacelike sections will still be given by a direct product
of $S^1 \times S^2$ (with constant $S^2$ radius), but the $S^1$ and
$S^2$ radii will not be equal. The flux runs around the $S^1$. The
metric is classically unstable to small perturbations.
Reissner-Nordstr\"om-de~Sitter black holes form where the two-sphere
size is smaller, and asymptotic de~Sitter regions develop where the
two-spheres are larger~\cite{BouHaw96,Bou99}.

These black holes are of nearly maximal size. For a small charge, one
would expect their initial behavior to be similar to the neutral
Schwarzschild-de~Sitter black holes.  When they are highly charged,
however, they become nearly extremal.  The temperature of a black hole
decreases as it approaches extremality.  Thermodynamic arguments were
given in Ref.~\cite{Bou99} which showed that there is a critical value
of the charge $Q_{\rm C}^2 = 3/(16 \Lambda) = 3/4 Q_{\rm
max}^2$. Roughly speaking, when a subcritical Nariai solution is
perturbed, the temperature of the black hole will be larger than that
of the cosmological horizon because it is smaller.  When a
supercritial Nariai solution is perturbed, the black hole will be
colder than the cosmological horizon, even though it is smaller,
simply because its mass is already very close to the mass of the
extremal, zero-temperature solution.

Thus one would expect subcritically charged black holes to evaporate
at late times. Supercritically charged black holes, on the other hand,
should absorb quantum radiation from the cosmological horizon and
grow, their size approaching the Nariai value asymptotically.  This is
confirmed numerically, as shown in Figs.~\ref{fig-z06Qw0per} and
\ref{fig-6zw0Q265per}. The value of the critical charge is confirmed
quantitatively.  This is a non-trivial check that this simple,
two-dimensional model reflects the thermodynamic properties of
Reissner-Nordstr\"om-de~Sitter black holes accurately.

The anti-evaporation of supercritially charged black holes has an
important consequence for the quantum global structure of de~Sitter
space.  It gives rise to a new type of proliferation effect, as
pointed out in Ref.~\cite{Bou99}.  Because the two-sphere size is
forever nearly constant in the Hubble-size region between the two
horizons, a small quantum fluctuation can easily invert the role of
the horizons.  In other words, it can increase the two-sphere size on
the black hole horizon, turning it into a cosmological horizon, and
vice-versa. This amounts to the insertion of a new black hole `bead'
into the $S^1$ `necklace.' Such processes repeat endlessly, so that an
unbounded number of causally disconnected de~Sitter regions develop on
the same $S^1$.

\section{Higher modes and proliferation}
\label{sec-proliferation}

\begin{figure}
\epsfxsize=0.45\textwidth 
\epsfbox{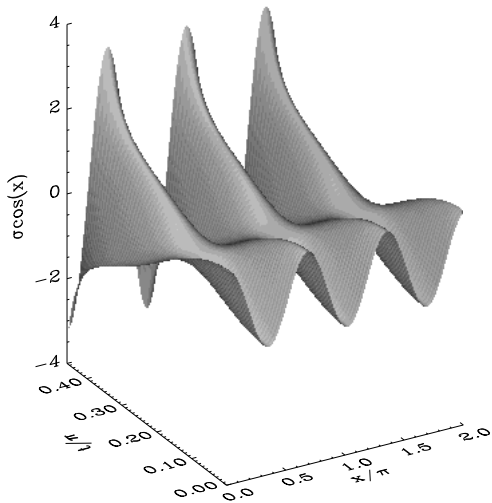}
\caption{\label{fig-zpi2w0n3surf} Metric perturbation evolution for
$n=3$ as function of $x$ and $t$. Model parameter values as in
Fig.~\ref{fig-6zw0per}.}
\end{figure}

We now return to uncharged solutions, but consider modes with $n>1$.
Such perturbations lead to a spatial geometry that can be described as
a `doughnut with $n$ wobbles.'  Like for $n=1$, the metric perturbation
is classically unstable.  The mode will oscillate until the $S^1$
expansion has stretched it enough to leave the horizon.  Then it will
freeze out, and grow exponentially.  This was shown in a linear
approximation in Ref.~\cite{Bou98}, and is demonstrated numerically in
Fig.~\ref{fig-zpi2w0n3surf} for $n=3$.

\begin{figure}
\begin{minipage}{0.45\textwidth}
\epsfxsize=\textwidth \epsfbox{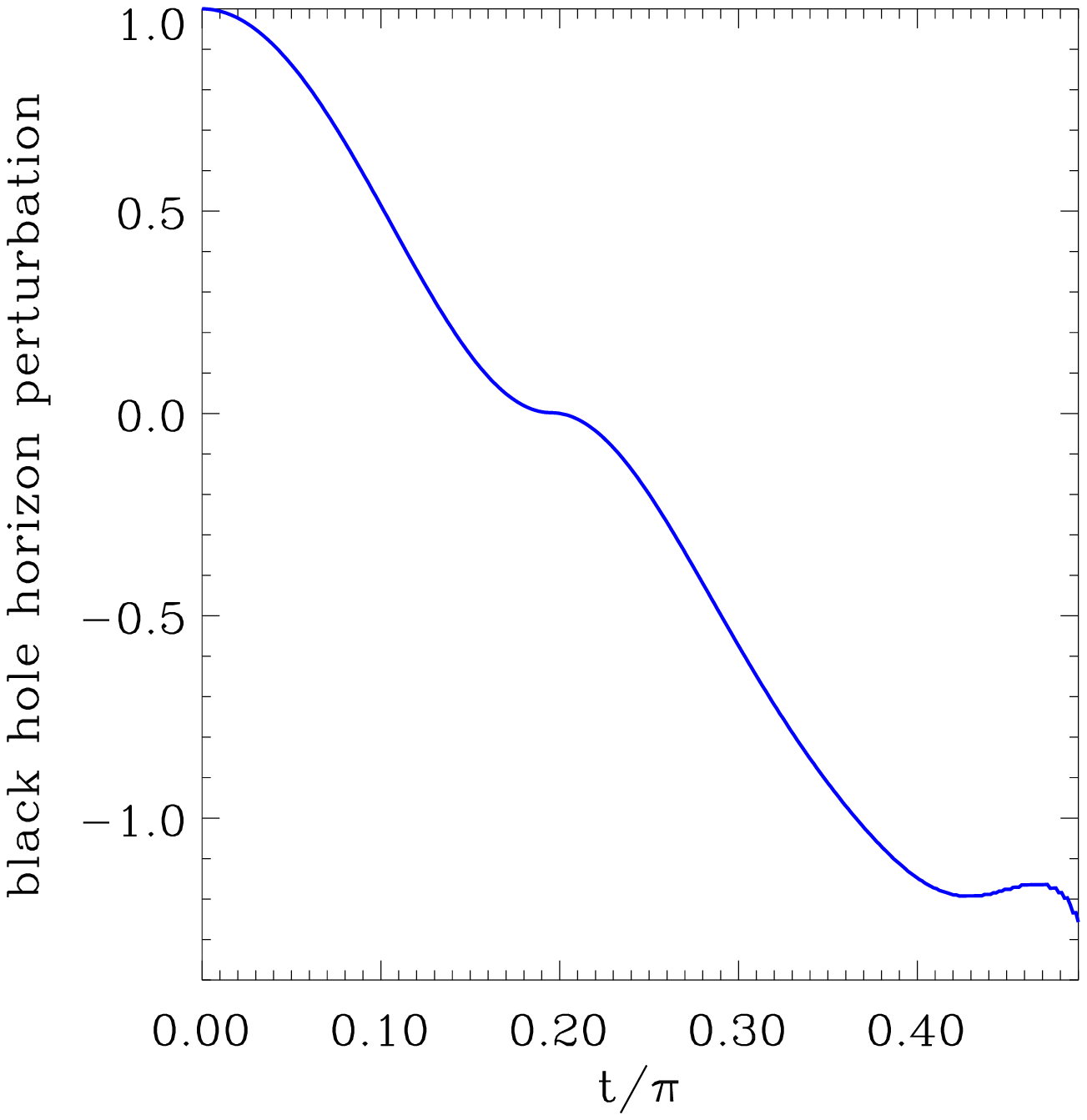}
\caption{\label{fig-zpi2w0n3per} Evolution of the black hole horizon
perturbation $\delta$ for the perturbation mode $n=3$. Model parameter
values as in Fig.~\ref{fig-6zw0per}.}
\end{minipage}
\hspace{\fill}
\begin{minipage}{0.45\textwidth}
\epsfxsize=\textwidth \epsfbox{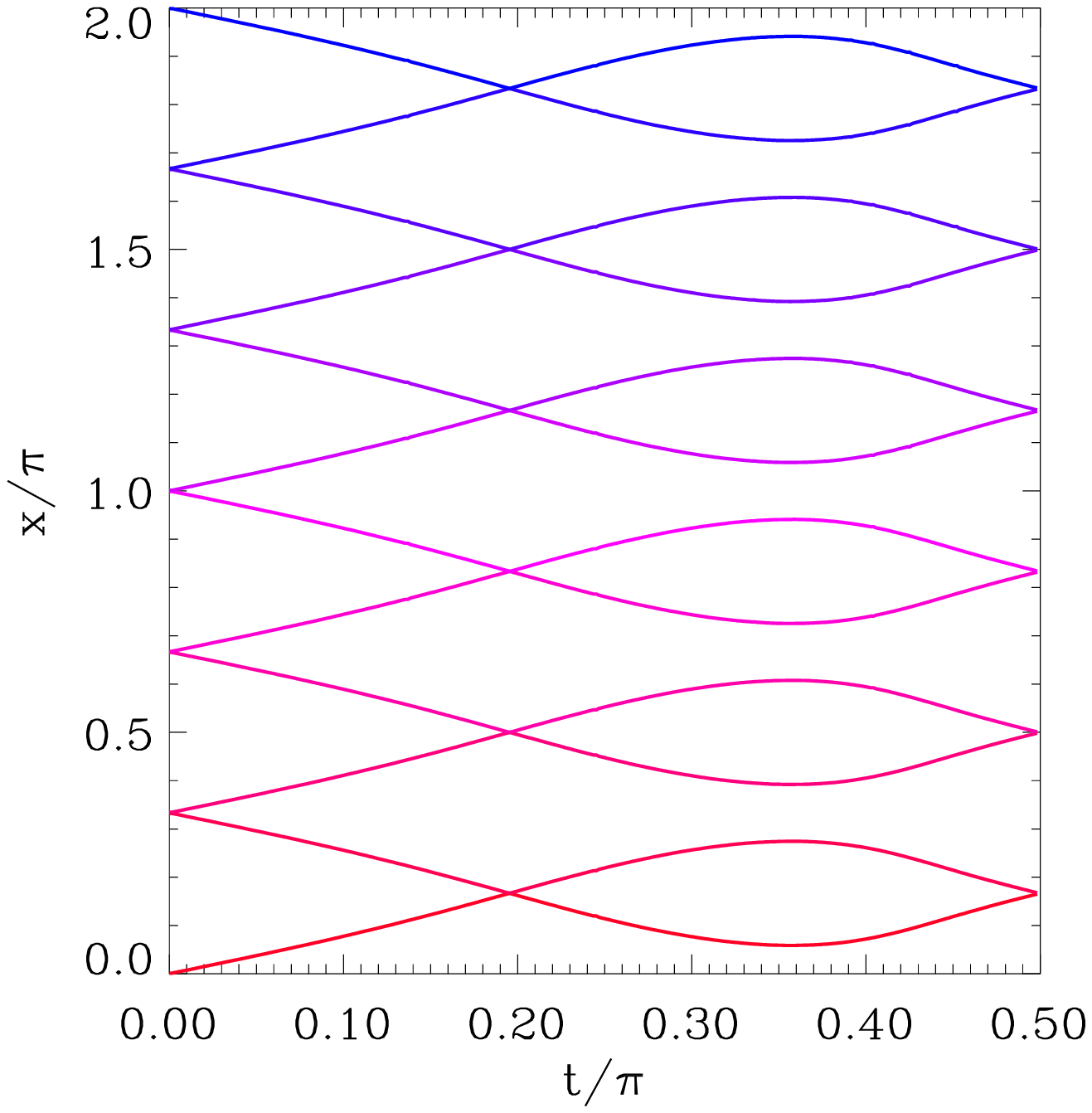}
\caption{\label{fig-zpi2w0n3hor} Location of the cosmological and
black hole horizons on the  $S^1$ coordinate for the perturbation mode
$n=3$.}
\end{minipage}
\end{figure}

The evolution of the horizon perturbation for $n=3$ is shown in
Fig.~\ref{fig-zpi2w0n3per}. While the metric perturbation oscillates,
the $(\nabla \phi)^2 = 0$ surfaces move rapidly and `cross over' (see
Fig.~\ref{fig-zpi2w0n3hor}).  They only represent black hole horizons
after the metric perturbation freezes out.  Fig.~\ref{fig-zpi2w0n3per}
shows that in the $n=3, \vartheta=\pi/2$ case considered there, the
metric perturbation freezes out while it of the opposite sign compared
to its initial value.  This means that cosmological horizons develop
from the initial minimal two-spheres.  This is reflected in the figure
in the negative values for the `black hole horizon' perturbation.
Note that the absolute value is increasing at late times, signalling
evaporation.

\section*{Acknowledgments}

We thank the Aspen Center for Physics where this
investigation was initiated, and the German American Academic Council
(GAAC) for financial support.

\end{document}